\documentclass{elsart}
\usepackage{epsfig}
\usepackage{amssymb,amsmath} 

\begin{document}
\newcommand{\DIR}{.}

\begin{frontmatter}          

\title{A Microscopic Model for Packet Transport in the Internet}

\author[Duisburg]{Torsten Huisinga}, 
\author[Duisburg]{Robert Barlovic}, 
\author[Duisburg]{Wolfgang Knospe}, 
\author[Cologne]{Andreas Schadschneider}, and 
\author[Duisburg]{Michael Schreckenberg}

\address[Duisburg]{Theoretische Physik, Gerhard-Mercator-Universit\"at 
Duisburg, Lotharstra{\ss}e 1, D-47048 Duisburg, Germany}

\address[Cologne]{Institut f\"ur Theoretische Physik, Universit\"at zu K\"oln,
  Z\"ulpicher Str.77, D-50937 K\"oln, Germany}

\begin{abstract}
A microscopic description of packet transport in the Internet by using
a simple cellular automaton model is presented. A generalised
exclusion process is introduced which allows to study travel times of
the particles ('data packets') along a fixed path in the
network. Computer simulations reveal the appearance of a free flow and
a jammed phase separated by a (critical) transition regime. The power
spectra are compared to empirical data for the RTT (Round Trip Time)
obtained from measurements in the Internet. We find that the model is
able to reproduce the characteristic statistical behaviour in
agreement with the empirical data for both phases (free flow and
congested). The phases are therefore jamming properties and not
related to the structure of the network. Moreover the model shows, as
observed in reality, critical behaviour ($1/f$-noise) for paths with
critical load.
\end{abstract}
\end{frontmatter}

\section{Introduction}

In recent years the Internet has become the most popular medium for
information transfer in the world. Terms like `e-mail' and
`e-commerce' are nowadays well known to almost everybody. Due to the
enormous increase of Internet users and a still growing demand the
network already reaches its maximum capacity at some times. Almost
every user has been annoyed by decreasing transfer rates and
increasing waiting times caused by congestions in the Internet. The
heterogeneity of the network, e.g., due to different transport
protocols and operating systems, and its enormous expansion in the
last years make it necessary to understand the basic properties of
data transport in the Internet for planning new connections and
optimising the usage of the existing resources. Especially the
influence of routers (network nodes) with low transfer rates, which
are considered to be the reason for the congestions, and the
collective behaviour of routers are main targets of recent
investigations. Real data measurements like those for the ping
statistics \cite{Csabai,Takayasu-1,Takayasu-2,huisi} or the load of a
single router \cite{Leland,Willinger-1} on various kinds of networks
and their analysis are the basis for a better understanding of
Internet traffic. Moreover there are investigations by Huberman and
Lukose \cite{huberman} on the social aspects of the Internet and the
``human factor'' in the system. Empirical results for the load of
single routers show a self-similar behaviour of Internet traffic which
Willinger {\it{et al.}} \cite{Willinger-1} explained as a
superposition of ON/OFF sources with heavy tailed distributions of the
duration lengths of the ON/OFF-periods. Another method to characterise
a nonequilibrium system like an Internet connection is the survey of
ping time series, first presented by Csabai \cite{Csabai} and later by
Takayasu {\it{et al.}}  \cite{Takayasu-1,Takayasu-3}. Here the travel
times of data packets from a source to a destination host and back to
the source host, the so-called Round Trip Times (RTT), are
measured. The analysis of the respective power spectra shows
characteristic statistics for different ``traffic'' states. One can
distinguish a free flow and a jammed phase separated by a transition
regime. On the basis of these measurements various models were
introduced to reproduce the characteristic stochastic
properties. Takayasu {\it{et al.}} \cite{Takayasu-2} proposed a simple
model based on the contact process \cite{Contact} to explain
$1/f$-noise in the travel times of data packets and to reproduce the
distribution of the congestion duration length of routers. The model
of Yuan {\it{et al.}}  \cite{yuan} is based on a reinterpretation of
the well-known cellular automaton approach for vehicular traffic
\cite{schreck-1,review}. Data transport is realised by changing
headways between ``moving routers''. This method does not give any
access to the travel times of data packets. In \cite{ohira} a
two-dimensional model has been suggested. Measurements of the travel
times indicate the existence of a phase transition into a jammed
phase. The influence of the structure of the network, namely the
branching number, has been investigated in \cite{vande} for a simple
stochastic model on a Cayley tree.

\section{Model}
\begin{figure}[!hbt] 
\begin{center} 
\epsfxsize=0.8\columnwidth\epsfbox{\DIR/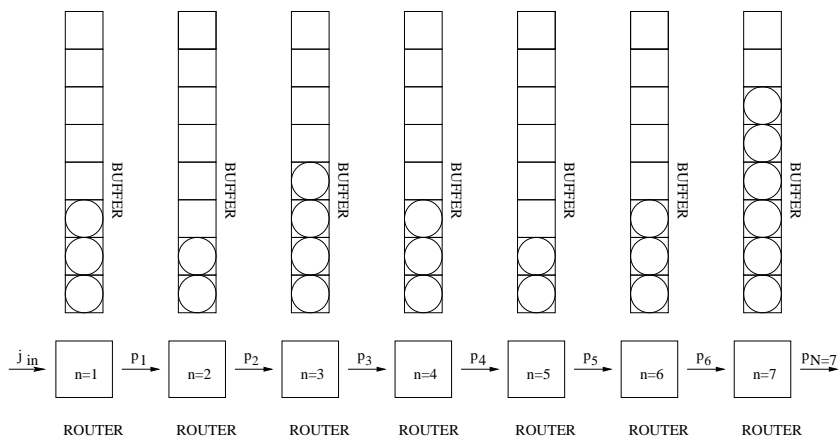} 
\end{center} 
\vspace{0.2cm} 
\caption{System consisting of $N=7$ routers with buffer size $B=8$.}
\label{model} 
\end{figure}

In the Internet traffic data files are divided into small data packets
of a definite size. These data packets move, for fixed source and
destination hosts, due to the structure of the Internet transportation
protocol (TCP/IP), along a temporally fixed route.  Therefore the
transport between two specific hosts can be viewed as a
one-dimensional process. Here we want to investigate the question
which properties of Internet traffic can already be understood by
considering just the one-dimensionality of these routes, i.e., as
jamming properties of the routers. A well known cellular automaton
model to describe one-dimensional transportation systems from
different fields like the kinetics of biopolymerisation and vehicular
traffic is the Asymmetric Simple Exclusion Process (ASEP)
\cite{review,ASEP}. Because of its simple structure it is a well
studied nonequilibrium system.  An important property of the ASEP is
the occurrence of boundary-induced phase transitions
\cite{krug}. Depending on the inflow and outflow the system can be in
different phases separated by (bulk) phase transitions. In order to
reproduce the statistical characteristics of Internet traffic we
introduce a simple microscopic cellular automaton model with open
boundary conditions based on the ASEP by allowing a finite number
$B_n$ of particles (data packets) on each site (router) $n$. Hereby we
take into account that each router has a buffer of finite capacity so
that more than one data packet can be stored (multi-allocation of
sites). The data packets move with a router specific probability $p_n$
to the the next router. This probability determines the amount of
traffic at the network node (the current) as well as the statistical
behaviour of processing times. The dynamics of the system do not only
depend on the probability a data packet moves to the next router, but
also on the restriction of the buffers so that a data packet only
moves to the next router as far as there is enough space left.\\ The
model is defined on a linear array of $N$ sites
(Fig.~\ref{model}). Each site $n=1,\ldots, N$ represents a router with
a buffer which stores $Z_n(t)$ particles at time $t$. Each router has
a finite capacity $B_n$, i.e., $Z_n(t) \leq B_n$. A particle $i$,
representing a data packet, moves with probability $p_n$ from site $n$
to the next site $n+1$ as long as the buffer ${n+1}$ is not completely
occupied. $p_n$ is a time-independent characteristic property of the
router $n$, i.e., it does not depend on the load of the buffer
itself. The update is performed in parallel for all buffers and the
travel times $T_i$ of all packets in the system are increased by the
discrete time $\Delta t$. The data packets arriving at the last site
$N$ are removed with probability $p_N$ and their travel times $T_i$,
i.e., the times needed to travel through the system, are stored.

\vspace{0.3cm}
  
\newpage
At $t=0$ we start with empty buffers at all routers, i.e.,
$Z_n(t=0)=0$. In each time step the following update steps are
applied in parallel:
\begin{enumerate}
\item
As long as the first router $n=1$ is not completely occupied
$j_{\text{in}}$ data packets are inserted: 
$Z_1(t+1)=\min(Z_1(t)+j_{\text{in}},B_1)$.
The travel times of these packets are set to zero: $T_i=0$.
\item
The travel times $T_i$ of those data packets $i$ present in the
system are increased by $\Delta t =1$.
\item
At each router $n=1,\ldots,N-1$ the data packets are picked up
sequentially in the order of their arrival in the buffer and move
with probability $p_n$ to the next router $n+1$ as long as this
router is not completely occupied ($Z_{n+1}(t)<B_{n+1}$).
\item
Data packets in the last router which have not already been moved in
the same time step are removed with probability $p_N$ and their travel
times $T_i$ are stored.
\end{enumerate}
Note that data packets in a buffer are stored in a waiting queue and
therefore the packets with the highest waiting times in the buffer
(not to be confused with the travel time) try to move first. Moreover
it is to mention that, due to the stochastic character of the movement
and the multi occupation of sites, particles can overtake each other
which can not be found in the ASEP.  Because of the parallel update
each data packet can move only once during each time step. In contrast
to \cite{vande} no data packets are lost. If $B_n=1$ for all $n$ the
model is identical to the ASEP (with disorder in the hopping rates)
with boundary probabilities $\alpha=1$ and $\beta=p_N$.

\section{Simulations}

\begin{figure}[hbt] 
\begin{center} 
\epsfxsize=0.5\columnwidth\epsfbox{\DIR/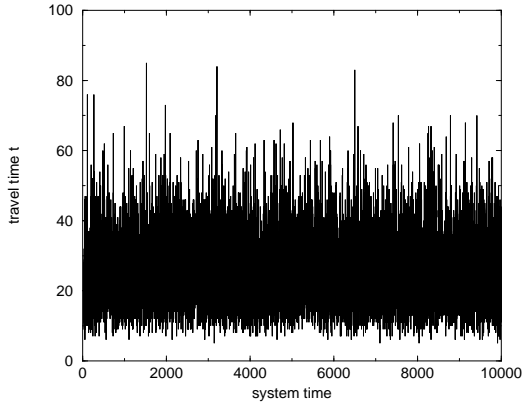} 
\end{center} 
\vspace{0.2cm} 
\caption{Typical sequence of travel times in a system of $N=5$ routers
with buffer size $B=128$, $p=0.2$, and $j_{in}=1$.}
\label{RTT-SIM} 
\end{figure}

First we investigate the phenomenological behaviour of the model. A
typical sequence of travel times is shown in Fig.~\ref{RTT-SIM}. Here
the travel times of the data packets are plotted as function of the
system time, i.e., the time at which the packet arrives at the end of
the system.

For the following investigations all routers have an identical buffer
size $B=128$. Note that with regard to reality we restrict the
number of routers to $N=15$ and the probabilities $p_{n}$ are
chosen in such a way to obtain a good agreement with empirical data.

As in the ASEP \cite{ASEP} the state of the system is determined by
the smallest of three currents, namely the maximal possible inflow,
bulk flow and outflow. The mean maximal flow $j^{\text{max}}_{n}$
through a single router $n$ is given by

\begin{equation}
j^{\text{max}}_{n}=B p_n. 
\end{equation}

For $j^{\text{in}}_{\text{n}} < j^{\text{max}}_{n}$ the dynamics of
the system is governed by the dynamics of the collective behaviour of
the routers. In reality congestions occur, when the amount of traffic
at a router exceeds its maximum capacity. In order to observe
congestion in the simulation we insert one single slow router with the
same capacity $B_{\text{def}}=B$, but with a lower moving probability
$p_{\text{def}}$. This router then behaves like a bottleneck
restricting the mean maximum flow to $j_{\text{def}}=B
p_{\text{def}}$. Because there is no major influence of the
unrestricted routers behind the bottleneck on the statistics of travel
times in the system, we associate the bottleneck of a path with the
boundary condition at the right end, i.e., $p_N=p_{\text{def}}$.
Since we are mainly interested in the impact of the slow router we
restrict the inflow $j_{\text{in}}$ to one data packet per update.\\

\begin{figure}[!hbt]
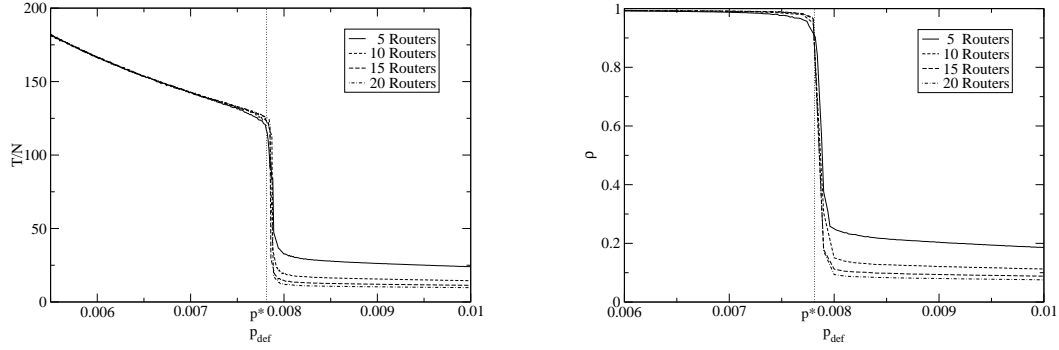

\centerline{\psfig{figure=\DIR/figure3.eps,width=6.3cm}\hfill \psfig{figure=\DIR/figure4.eps,width=6.3cm}}
\caption{\protect{Left: Diagram of the mean travel times $T$ of data
packets versus the probability $p_{\text{def}}$ of the last router for
($N=5,10,15,20$), $B=128$, and $p_n=0.2$. The mean travel times have
been sampled over 500.000 updates after relaxation into the steady
state. Right: Relation between the density $\rho$ of data packets and
the probability $p_{\text{def}}$ of the last router. The parameters
correspond to the ones used for the left diagram.}}
\label{p-t}
\end{figure}   

Varying $p_{\text{def}}$, computer simulations reveal the existence of
two phases which can be distinguished by the behaviour of the travel
times and the average density in the system (Fig.~\ref{p-t}). The
travel times are obtained from the simulations by summing up the waiting
times $\tau_{i,n}$ of every single data packet $i$ in the routers
along the path: $T_i=\sum_{n=1}^N \tau_{i,n}$. For a free flow system
the mean waiting time for an arbitrary data packet in router $n$ can
be estimated by
\begin{equation}
\tau_{n} =\sum_{t=0}^{\infty}tp_n(1-p_n)^{t-1}=\frac{1}{p_n}.
\label{t-free}
\end{equation}
The behaviour of the system is determined by the relation between
$p_{\text{def}}$ and $p^{*}$ where $p^*$ corresponds to the point of
maximum bulk flow. A simple estimation for $p^*$ is
\begin{equation}
p^* = \frac{j_{\text{in}}}{B_{\text{def}}} = \frac{1}{B},
\label{p*}
\end{equation}   
i.e., for $p_{\text{def}}=p^*$ the inflow is equal to the mean maximum
flow $p_{\text{def}}B_{\text{def}}$ through the last router.

For moving probabilities $p_{\text{def}} > p^{*}$ the maximum flow
through the bottleneck is higher than the inflow $j_{\text{in}}$. In
this free flow system the mean travel time only depends on the average
capacity of each single router and can be described by
\begin{equation}
T_{\text{free}} = \sum_{n=1}^{N} \tau_{n} = \sum_{n=1}^{N} \frac{1}{p_{n}}
=\sum_{n=1}^{N-1} \frac{1}{p_{n}}+\frac{1}{p_{\text{def}}}.
\label{t-mean-free}
\end{equation}

For lower moving probabilities ($p_{\text{def}} < p^{*}$) the mean
flow through the bottleneck is lower than the inflow $j_{\text{in}}$
and the system gets jammed. In the jammed state, the maximum system
flow is determined by the maximum capacity of the bottleneck. Data
packets can only move to the next router when a data packet left it a
time step before. This means that the mean travel time through the
system is given by
\begin{equation}
T_{\text{jam}} = \frac{N}{p_{\text{def}}}.
\label{t-mean-jam}
\end{equation}

For $p_{\text{def}} \approx p^{*}$ the mean flow through the
bottleneck is equal to the inflow $j_{\text{in}}$ which means that the
system operates at its maximum capacity. Simulations show that
(\ref{t-mean-free}) and (\ref{t-mean-jam}) are in excellent agreement
with the results from Fig.~\ref{p-t}.

The existence of two well defined regimes in the presence of a defect
router $p_{\text{def}}$ is also confirmed by measurements of the mean
density $\rho$ of the system (see Fig.~\ref{p-t}) which is defined by
$\rho = \sum_{n=1}^{N} \frac{Z_n}{B}$. In Fig.~\ref{p-t} one can
distinguish a free flow state with low density for $p_{\text{def}} >
p^{*}$ and a jammed state with high density for $p_{\text{def}} <
p^{*}$ in agreement with the results for the travel time.

\begin{figure}[!hbt]
\centerline{\psfig{figure=\DIR/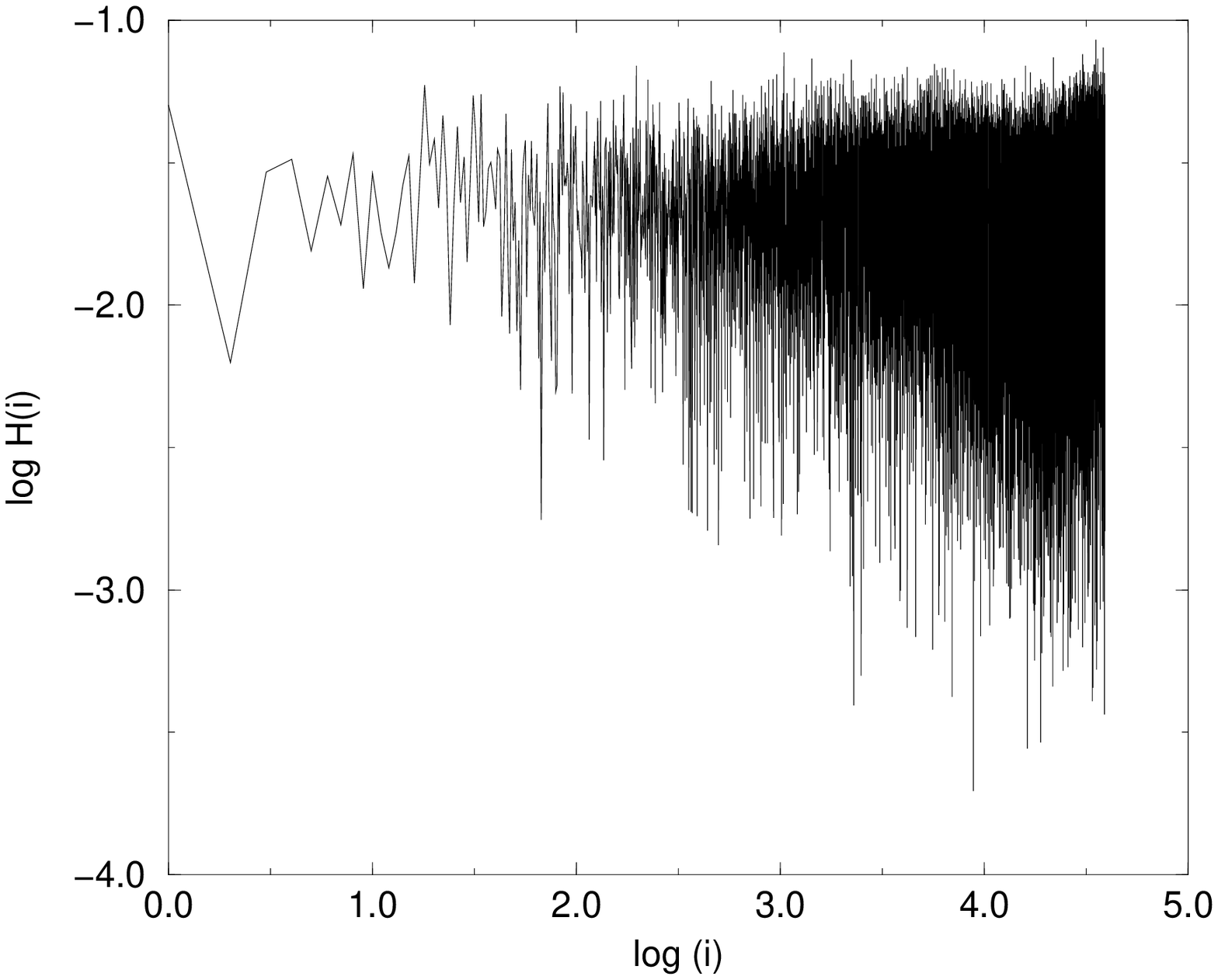,width=6.3cm}\hfill \psfig{figure=\DIR/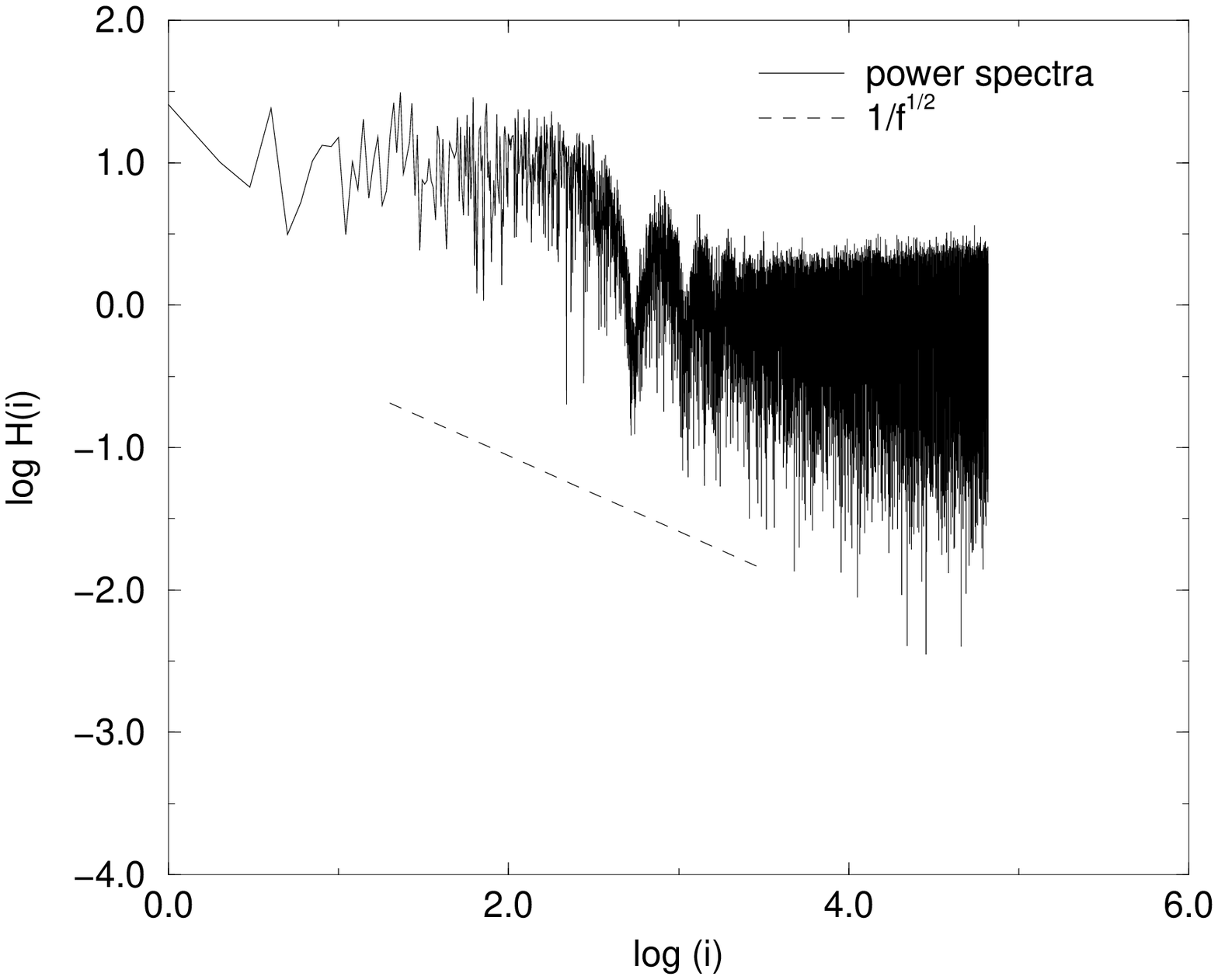,width=6.3cm}}
\caption{\protect{ Left: Power spectrum of a free flow system
showing white noise ($N=15$, $B=128$, $p=0.2$, and
$p_{\text{def}}=0.1$). Right: Power spectrum of a jammed system
with $1/f^{1/2}$-noise at low and white noise at high frequencies
($N=15$, $B=128$, $p=0.2$, and $p_{\text{def}}=0.003$). }}
\label{LS-free}
\end{figure}

\begin{figure}[!hbt]
\begin{center}
\epsfxsize=0.5\columnwidth\epsfbox{\DIR/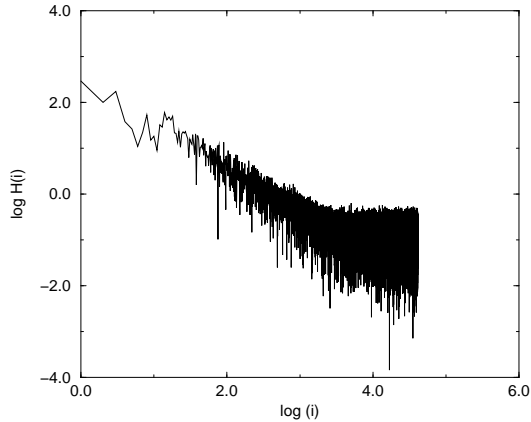}
\end{center}
\vspace{0.2cm}
\caption{Power spectrum of a system at critical load ($N=15$,
$B=128$, $p=0.2$, and $p_{\text{def}}=0.788$). One finds $1/f$-noise 
at low frequencies and white noise at high frequencies.}
\label{LS-crit} 
\end{figure}

To compare the simulation results with empirical travel times from
ping experiments, we investigated the statistics in the jammed and the
free flow regime as well as in the transition region between these two
regimes at $p_{\text{def}}=p^*$. Therefore we generated the power
spectra of the travel times and analysed the spectral density.  The
left part of Fig.~\ref{LS-free} shows the power spectrum of a free
flow system ($p_{\text{def}} \gg p^{*} = 1/B$). White noise is found
for the whole frequency range. This means that correlations in the
travel times of the data packets are negligible small. The data
packets move with probability $p_n$ from one router to the next one
without any limitation caused by the buffer restrictions. In contrast,
jammed systems ($p_{\text{def}} \ll p^{*}$) show a algebraic decay
with an approximately $1/f^{1/2}$ dependence at low frequencies (see
right part of Fig.~\ref{LS-free}). Considering the occupancy of the
buffer as a time dependent variable, the interval distribution of one
jammed buffer corresponds to the first recurrence time in the random
walk problem. Such a system then shows $1/f^{1/2}$-noise in the power
spectrum and white noise at higher frequencies \cite{Takayasu-1}. In
the transition regime in the vicinity of $p^*$ the power spectra of
the travel times show characteristic $1/f$-noise (see
Fig.~\ref{LS-crit}) at low frequencies (long range correlations,
critical behaviour). All of the above findings of the statistical
analysis of travel times generated by simulations of our simple model
are in full agreement with the characteristic properties of
measurements of ping time series in the Internet
\cite{Takayasu-1,huisi}.

\section{Discussion}

We have introduced a simple cellular automaton model for the Internet
data packet transport along a fixed path in the Internet. It is an
asymmetric exclusion process where occupation of sites (buffers) by
more than one particle (data packet) is allowed. Computer simulations
have revealed the occurrence of a jammed and free flow phase in the
presence of a slow router. To compare our model with real Internet
data we focused on the dynamic behaviour of the travel times and their
correlations. The analysis of travel times shows the typical power
spectra of real Internet traffic in the two regimes, i.e., white noise
for free flow and $1/f^{1/2}$ for the jammed system. In the transition
regime between these two phases the model shows a characteristic
$1/f$-noise.\\ In this work we focused on the effects due to one slow
router in a fixed packet transport path, i.e., $p_{\text{def}}$.  The
influence of other parameters, e.g., $j_{\text{in}}$,
$B_{\text{def}}$, $p_n$, etc., has been investigated in
\cite{huisi}. The results will be reported elsewhere. Future work
should characterise the transition in more detail. In order to
simulate the behaviour of networks where the nodes act as source and
destination hosts the model has to be extended to two
dimensions. However, our investigations indicate that many of the
statistical properties of Internet traffic can already be understood
by the simple one-dimensional model.

\end{document}